\documentclass[aps,pre,onecolumn,notitlepage,superscriptaddress,groupedaddress]{revtex4}

\usepackage{xcolor,graphicx}
\usepackage{amssymb,amsmath,graphicx}
\usepackage[utf8]{inputenc}
\usepackage{comment}
\usepackage{braket}
\newcommand{\me}[1]{\left\langle #1 \right\rangle }

\begin{document}

\title{Energy fluctuation relations and repeated quantum measurements}

\author{Stefano Gherardini}
\email{stefano.gherardini@ino.cnr.it}
\affiliation{CNR-INO, Area Science Park, Basovizza, I-34149 Trieste, Italy}
\affiliation{SISSA, via Bonomea 265, I-34136 Trieste, \& INFN, Italy}
\affiliation{LENS, University of Florence, via G. Sansone 1, I-50019 Sesto Fiorentino, Italy}

\author{Lorenzo Buffoni}
\email{lbuffoni@lx.it.pt}
\affiliation{Physics of Information and Quantum Technologies Group, Instituto de Telecomunicaç\~oes, University of Lisbon, Av. Rovisco Pais, P-1049-001 Lisbon, Portugal}

\author{Guido Giachetti}
\email{ggiachet@sissa.it}
\affiliation{SISSA and INFN, Sezione di Trieste, Via Bonomea 265, I-34136 Trieste, \& INFN, Italy}

\author{Andrea Trombettoni}
\email{andreatr@sissa.it}
\affiliation{Department of Physics, University of Trieste, Strada Costiera 11, I-34151 Trieste, Italy}
\affiliation{SISSA and INFN, Sezione di Trieste, Via Bonomea 265, I-34136 Trieste, \& INFN, Italy}
\affiliation{CNR-IOM DEMOCRITOS Simulation Center, Via Bonomea 265, I-34136 Trieste, Italy}

\author{Stefano Ruffo}
\email{ruffo@sissa.it}
\affiliation{SISSA and INFN, Sezione di Trieste, Via Bonomea 265, I-34136 Trieste, \& INFN, Italy}
\affiliation{Istituto dei Sistemi Complessi, Consiglio Nazionale delle Ricerche, via Madonna del Piano 10, I-50019 Sesto Fiorentino, Italy}

\begin{abstract}
In this review paper, we discuss the statistical description in non-equilibrium regimes of energy fluctuations originated by the interaction between a quantum system and a measurement apparatus applying a sequence of repeated quantum measurements. To properly quantify the information about energy fluctuations, both the exchanged heat probability density function and the corresponding characteristic function are derived and interpreted. Then, we discuss the conditions allowing for the validity of the fluctuation theorem in Jarzynski form $\langle e^{-\beta Q}\rangle = 1$, thus showing that the fluctuation relation is robust against the presence of randomness in the time intervals between measurements. Moreover, also the late-time, asymptotic properties of the heat characteristic function are analyzed, in the thermodynamic limit of many intermediate quantum measurements. In such a limit, the quantum system tends to the maximally mixed state (thus corresponding to a thermal state with infinite temperature) unless the system's Hamiltonian and the intermediate measurement observable share a common invariant subspace. Then, in this context, we also discuss how energy fluctuation relations change when the system operates in the quantum Zeno regime. Finally, the theoretical results are illustrated for the special cases of two- and three-levels quantum systems, now ubiquitous for quantum applications and technologies.
\end{abstract}

\maketitle

\section{Introduction}

The study of fluctuation relations covers several fields of physics, from statistical physics to quantum mechanics and quantum optics. Tito Arecchi gave seminal and inspiring contributions to this field, beginning with his pioneering works on photon statistics \cite{Arecchi1972} and photon fluctuation in lasers near bifurcation points \cite{Arecchi1982}. This latter paper was among the first ones verifying the Feigenbaum bifurcation cascade to chaos. His curiosity for fundamental problems accompanied him through all his scientific life, including a long-lasting interest for quantum mechanics and the subtleties of the quantum measurement. One of us, SR, remembers with pleasure several discussions with Tito on Bell's inequality, a subject of his Master Thesis.

At the microscopic level, matter is in a permanent agitation state; consequently, many physical quantities of interest undergo random fluctuations. In this regard, one of the objectives of statistical mechanics is to describe the properties of these fluctuating quantities, using the laws of classical and quantum physics \cite{ZwanzigBook,McQuarrieBook}. A standard example of this concept is Maxwell's velocity distribution of the particles contained in an ideal gas at thermodynamic equilibrium, whose statistical property can be derived just using the law of large numbers and fundamental symmetries. Moreover, quantities such as work, heat and entropy in any non-equilibrium transformation are subject to fluctuations. The relations between these quantities, provided by classical thermodynamics, are in the form of inequalities \cite{deGrootBook}.

In the past twenty years, several studies have yielded significant advances in the field of non-equilibrium dynamics and thermodynamics \cite{Esposito2009,Campisi2011}. Indeed, it has been understood that fluctuations of quantities such as work, heat and entropy satisfy properties that, if correctly treated, allow to re-express the standard inequalities of macroscopic thermodynamics, such as $\langle W\rangle \geq \Delta F$ where $\langle W\rangle$ is the average work done by the system and $\Delta F$ the corresponding free energy variation, as equalities. The first work that gave the start to this period of great developments in statistical mechanics is the one by Gallavotti and Cohen \cite{GallavottiPRL1995}. From this paper new results followed, the most important of which are denoted as \emph{fluctuation theorems}, respectively, by Jarzynski \cite{JarzynskiPRL1997} and Crooks \cite{CrooksPRE1999}, respectively in 1997 and 1999. Jarzynski and Crooks fluctuation theorems relate the statistics of the work done by an external field on the system of interest, in a finite time interval through a non-equilibrium process, with the free-energy variation that can be measured only when the system is at equilibrium. Originally derived for classical systems, fluctuation theorems have been transposed to the quantum case by Kurchan in 2001 \cite{Kurchan2001} and Mukamel in 2003 \cite{MukamelPRL2003}. After that, several other contributions have addressed the problem of appropriately describing energy fluctuation relations for open quantum systems \cite{CampisiPRL2009,Albash13PRE88,FuscoPRX2014,Campisi15NJP17,GherardiniEntropy,ManzanoPRX2018,CiminiNPJ2020} and especially systems subject to a sequence of repeated quantum measurements, both projective and generalized ones \cite{Campisi2010PRL,Campisi2011PRE,Yi2013,WatanabePRE2014,WatanabePRE2014,Elouard2016,GarciaPintosPRL2019,MartinsPRA2019}. Overall, in the quantum systems case, it has been determined that the fluctuations theorems still remain valid provided that (i) initially the system is at thermodynamic equilibrium \cite{TalknerPRE2007}; (ii) the system's evolution is governed by a unital quantum map, whereby the identity is a fixed point of the map \cite{Rastegin13JSTAT13,RasteginPRE2014}. Moreover, it is worth also mentioning results and applications \cite{Gherardini2016NJP,Mueller2017ADP,Gherardini2017QSc,GherardiniPRR2020Batteries} concerning the quantum Zeno effect \cite{BurgarthQU2020}, and the works by Sagawa and Ueda \cite{SagawaPRL2008,SagawaPRL2010}, together with related studies \cite{ParrondoNatPhys2015,MasuyamaNatComm2018,Campisi17NJP19,HernandezArXiv2021}, that deal with the informational aspect of non-equilibrium thermodynamics by studying the dynamics of a quantum system subject to closed-loop control. Finally, protocols involving only series of quantum measurements or cyclical repetitions of them, have shown also the ability to fuel quantum heat engines \cite{buffoni2019quantum,bresque2021two}, provided that the right measurement observable is chosen \cite{solfanelli2019maximal}.
These findings open the possibility of using protocols of repeated quantum measurements not only as a theoretical tool to investigate the fundamental aspects of quantum thermodynamics but also as a practical way to fuel engines at the micro and nanoscale.

In this paper we discuss the interplay between energy fluctuation relations and repeated quantum measurements. Our presentation make use of results in \cite{GherardiniPRE2018} and \cite{GherardiniPRE2021}, and as well on few unpublished results presented in a Thesis of one of us \cite{TesiLorenzo}. Both papers \cite{GherardiniPRE2018,GherardiniPRE2021} concern the statistics of the energy exchanged between a discrete quantum system and its surrounding (e.g., provided by the external measurement apparatus), but our presentation is new since our aim is to present to the general audience a comprehensive formalism in which the different cases scrutinized in \cite{GherardiniPRE2018,GherardiniPRE2021} are conveniently retrieved. We will discuss the fact that the Jarzynski equality of the exchanged energy is obeyed independently on the fact that is present randomness in the distribution of the waiting times $\tau$ between consecutive measurements. Experimentally, this result has been confirmed in \cite{HernandezPRR2020} where a nitrogen-vacancy center spin qubit in diamond at room temperature has been employed. The second result pertains the asymptotic behaviour of the exchanged energy distribution in the limit of large number of measurements. Moreover, further unusual effects have been determined in the limiting case of an ideally infinite number of levels (thus approaching the continuous limit). Specifically, one can analytically prove that, if the Hamiltonian $H$ of the quantum system and the observable $\mathcal{O}$ of the intermediate measurements do not share any common non-trivial subspace, then the system asymptotically thermalizes to a thermal state with infinite temperature, as also argued in Refs.\,\cite{GurvizPRL2000,YiPRA2011}. This effect has a clear physical interpretation, since it necessarily entails the complete mixing of the eigenvectors of the intermediate observable $\mathcal{O}$ at the end of the non-equilibrium quantum process. Instead, the infinite-temperature thermalization is hindered (partial thermalization) whenever $H$ and $\mathcal{O}$ have one or more eigenvectors in common (e.g. when $[H,\mathcal{O}]=0$), again independently of the waiting times distribution. Furthermore, partial thermalization can occur when the value of the waiting times is on average much smaller than the inverse of the energy scale of the system (phenomenon ascribable to the quantum Zeno effect), and also in the large-$N$ limit.

The paper is organized as follows. In Sec.\,II, we provide recent theoretical results obtained for a generic (discrete-variable) quantum system subject to a sequence of repeated projective measurements. Specifically, after having introduced the general formalism of the applied protocol, we define the probability distribution and characteristic function of the energy exchanged by the system as an effect of its monitoring. Then, we discuss the conditions that allow for the validity of a fluctuation theorem for the exchange energy statistics, in the form of the celebrated Jarzynski identity. Furthermore, the asymptotic behaviour of a generic monitored quantum system in the limit of many (ideally infinite) intermediate measurements is also discussed. In this way, the quantum Zeno effect will be here introduced as an exception of the regime denoted as infinite-temperature thermalization occurring asymptotically. In Sec.\,III, instead, explicit calculations and simulations on two- and three-level quantum systems are shown as illustrative examples of the general theoretical analysis. Finally, Sec.\,IV is devoted to our conclusions.

\section{Protocols of repeated quantum measurements}

In this section, first we introduce the general formalism describing the effect on applying a sequence of repeated projective measurements on a (discrete-variable) quantum system, and then we present some important results related to this topic -- both at the theoretical and experimental sides -- that some of us have been recently derived in Refs.\,\cite{GherardiniPRE2018,HernandezPRR2020,GiachettiCM2020,GherardiniPRE2021}.

\subsection{General formalism}

Let us consider a quantum system to which we associate a $N$-dimensional Hilbert space. The Hamiltonian of the system, that is denoted as $H$ and assumed to be time-independent, can be expressed according to the spectral decomposition, i.e.,
\begin{equation}
H = \sum^N_{k=1} E_k |E_k\rangle\!\langle E_k|\,.
\end{equation}
with $\{E_k\}$ set of system energies and $\{|E_k\rangle\}$ denoting the set of corresponding energy eigenvectors. Note that, in using such decomposition, no assumption about the degeneracy of the eigenstates of $H$ has to be taken into account.

Initially, at time $t=0^{-}$, the quantum system is in an arbitrary quantum state described by the density operator $\rho_{0^{-}}$, while at time $t = 0^{+}$ a first energy projective measurement is performed on the system with the result that its state collapse -- instantaneously for all practical purposes -- to one of the energy projectors, such that $\rho_{0^{+}} \equiv |E_n\rangle\!\langle E_n|$, with probability
$p_n \equiv {\rm Tr}[\rho_{0^{-}}|E_n\rangle\!\langle E_n|]$. The need to perform an initial energy projective measurement on the quantum system stems directly from the use of the so-called Two-Point Measurement (TPM) scheme \cite{TalknerPRE2007,Kafri2012}. According to such scheme, to the quantum system can be assigned as initial energy value the eigenvalue $E_n$ associated to the energy projector $|E_n\rangle\!\langle E_n|$ on which the system has been made collapse. If on one hand, applying the TPM scheme has thus the advantage to remove any ambiguity on the initial system's energy, it has the drawback to do not take into account the possible presence of initial quantum coherence/correlation terms in the initial density operator $\rho_{0^{-}}$ that, to all the effects, can be simply taken diagonal with respect to the system's energy basis (spanned by the eigenstates of $H$), i.e., $\rho_{0^{-}} = \sum^N_{k=1} c_k |E_k\rangle\!\langle E_k|$ with $c_k >0$ $\forall k = 1,\dots,N$ and $\sum^N_{k=1} c_k = 1$. Recently, efforts to overcome this issue have been already made \cite{Solinas2015,SonePRL2020,MicadeiPRL2020,LevyPRXQ2020,GherardiniarXiv2021}, but none of these solutions have been applied to the case of quantum protocols of repeated measurements. Here, we are going to present results obtained by applying the TPM scheme, but we will precisely point out the assumptions that one could remove to also take into account the presence of quantum coherence/correlation terms in the exchange energy statistics.

After the first energy projective measurement prescribed by the TPM scheme, the quantum system is subject to a number $M$ of consecutive projective measurements on the generic observable
\begin{equation}
\mathcal{O} \equiv \sum^N_{k=1} \alpha_k |\alpha_k\rangle\!\langle\alpha_k| \,,
\end{equation}
with $\alpha_k$ and $|\alpha_k\rangle$ denoting the outcomes and eigenstates of $\mathcal{O}$, respectively. According to the postulates of quantum mechanics, after each projective measurement the system collapses in a pure state provided by one of the projectors $|\alpha_k\rangle\!\langle\alpha_k|$. During the interval between the $(j-1)^{\text{th}}$ and the $j^{\text{th}}$ measurement of $\mathcal{O}$, the evolution of the system is unitary and governed by the operator $U(\tau_j) = e^{-iH\tau_j}$ where in the formula the reduced Planck’s constant $\hbar$ is set to unity and the waiting times $\tau_j$ are the interval between two consecutive measurements. As it will be shown below, the waiting times may not be deterministic quantities, but random variables following the joint probability density function $p(\tau_1,\dots,\tau_M)$. By using the vectors $\boldsymbol\tau \equiv (\tau_1,\dots,\tau_M)$ and $\mathbf{k} \equiv (k_1,\dots,k_M)$ to denote respectively the sequences of waiting times and outcomes obtained by measuring $\mathcal{O}$, the state of the system after the $M$-th measurement of $\mathcal{O}$ can be written as
\begin{equation}
\rho_M = \frac{\mathcal{V}_{\mathbf{k}, \boldsymbol\tau} \rho_{0^{+}}\mathcal{V}^{\dagger}_{\mathbf{k}, \boldsymbol\tau}}{{\rm Tr}[\mathcal{V}_{\mathbf{k}, \boldsymbol\tau}\rho_{0^{+}}\mathcal{V}^{\dagger}_{\mathbf{k}, \boldsymbol\tau}]}\,,
\end{equation}
where $\mathcal{V}_{\mathbf{k},\boldsymbol \tau} \equiv |\alpha_{k_M}\rangle\!\langle\alpha_{k_M}| U(\tau_M) \cdots |\alpha_{k_1}\rangle\!\langle\alpha_{k_1}| U(\tau_1)$ and $\rho_{0^{+}} \equiv |E_{n}\rangle\!\langle E_{n}|$. Once again, we point out that the value of $\rho_{0^{+}}$ stems from applying the TPM scheme to characterize energy statistics of the non-equilibrium process of interest. This means that, to evaluate the initial presence of quantum coherence/correlation terms, the initial state of the system shall not be taken as one of the energy projectors diagonalizing $H$ nor a composition of them.

In conclusion of the protocol, a second energy measurement is performed immediately after the last (i.e., the $M$-th) measurement of $\mathcal{O}$, such that $\rho_{M^{+}} = |E_m\rangle\!\langle E_m|$ is the final state of the system.

Finally, by calling the outcome of the second and final energy measurement as $E_m$, the internal system's energy variation $\Delta U$ is provided by the following difference among the stochastic realizations of the energy values recorded at the initial and final time instants of our protocol:
\begin{equation}\label{eq:def_Q}
\Delta U \equiv E_m - E_n \,.
\end{equation}
Since the performed projective measurements are random exogenous genuinely-quantum processes, we the internal energy variation $\Delta U$ is classified as heat $Q$ absorbed or emitted by the system \cite{GherardiniPRE2018,GherardiniPRE2021}.

\subsection{Exchanged energy statistics}

Let us provide the expressions of the exchanged energy distribution and characteristic function. In this regard, it is worth considering that the heat $Q$ is a random variable, due to (i) the probabilistic nature of each outcome in measuring $\mathcal{O}$ (contained in the sequence $\mathbf{k}$); (ii) a possible randomness in the way the waiting times are distributed in the sequence $\boldsymbol \tau$; (iii) initial statistical mixture, identified by the initial density operator $\rho_{0^{-}}$ that from here on will be simply denoted as $\rho_{0}$. Accordingly, the probability distribution of $Q$ reads
\begin{equation}\label{eq:pheat}
{\rm Prob}(Q) = \sum_{n,m}\delta(Q - E_{m} + E_{n})\,p_{m|n}\,p_{n}\,,
\end{equation}
with $p_{m|n}$ denoting the conditional probability to measure the final energy $E_{m}$ conditioned to have obtained $E_{n}$ from the first energy measurement required by the TPM scheme. If we denote with $p_{m|n}(\mathbf{k},\boldsymbol \tau)$ the probability to measure a transition from $n$ to $m$ conditioned on the outcomes and waiting times sequences $\mathbf{k},\boldsymbol \tau$ (also called \emph{conditioned transition probability}), then the conditional probability $p_{m|n}$ can be expressed as
\begin{equation}\label{conditional_prob_an}
p_{m|n} = \int\sum_{\mathbf{k}}d^{M}\boldsymbol \tau\,p_{m|n}(\mathbf{k},\boldsymbol \tau)\,p(\boldsymbol \tau)\,,
\end{equation}
where $p(\boldsymbol \tau) \equiv p(\tau_1,\dots,\tau_M)$ is the joint probability density function of $\boldsymbol \tau$ and the conditioned transition probability $p_{m|n}(\mathbf{k},\boldsymbol \tau)$ reads
\begin{eqnarray}
p_{m|n}(\mathbf{k},\boldsymbol \tau) &=& {\rm Tr}\left[|E_m\rangle\!\langle E_m|\mathcal{V}_{\mathbf{k},\boldsymbol \tau}|E_n\rangle\!\langle E_n|\mathcal{V}^{\dagger}_{\mathbf{k},\boldsymbol \tau}|E_m\rangle\!\langle E_m|\right]\nonumber \\
&=& \langle E_{m}|\mathcal{V}_{\mathbf{k},\boldsymbol \tau}|E_{n}\rangle\langle E_{n}|\mathcal{V}^{\dagger}_{\mathbf{k},\boldsymbol \tau}|E_{m}\rangle\,.
\end{eqnarray}

The derivation of the heat probability distribution ${\rm Prob}(Q)$ may go through the computation of the corresponding characteristic function
\begin{equation}\label{eq:G(u)}
G(u) \equiv \int {\rm Prob}(Q)e^{iuQ}dQ\,,
\end{equation}
with $u\in\mathbb{C}$ complex number. In fact, the characteristic function $G(u)$ can be evaluated by resorting to Ramsey interferometry schemes \cite{DornerPRL2013,MazzolaPRL2013,BatalhaoPRL2014,BatalhaoPRL2015}, or to methods from estimation theory \cite{GherardiniEntropy}. These experimental procedures becomes feasible by noting that, after substituting first Eq.\,(\ref{conditional_prob_an}) in Eq.\,(\ref{eq:pheat}) and then plugging Eq.\,(\ref{eq:pheat}) into (\ref{eq:G(u)}), $G(u)$ reads
\begin{equation*}
G(u) = \int d^{M}\boldsymbol \tau \, p(\boldsymbol \tau)\sum_{n,\mathbf{k},m}
\langle E_{m}|\mathcal{V}_{\mathbf{k},\boldsymbol \tau}|E_{n}\rangle\langle E_{n}|\rho_{0^{-}}|E_{n}\rangle\langle E_{n}|e^{-iuH}\mathcal{V}^{\dagger}_{\mathbf{k},\boldsymbol \tau}e^{iuH}|E_{m}\rangle \,,
\end{equation*}
where also the eigenvalue equations $e^{iuE_{m}}|E_{m}\rangle = e^{iuH}|E_{m}\rangle$ and $\langle E_{n}|e^{-iuE_{n}}=\langle E_{n}|e^{-iuH}$ have been exploited. In this way, by adopting the short-hand notation used in Ref.\,\cite{GherardiniPRE2018} whereby the expectation over all the possible measurements outcomes is denoted by the angular brackets [i.e., $\langle\cdot \rangle \equiv \sum_{\mathbf{k}}(\cdot)$] and the average over noise realisations by the overline [namely $\overline{(\cdot)} \equiv \int d^M\boldsymbol \tau\,p(\boldsymbol \tau)(\cdot)$], the final expression of $G(u)$ is derived:
\begin{equation}\label{G_u_finale}
G(u) = \overline{\left\langle {\rm Tr}\left[e^{iuH}\mathcal{V}_{\mathbf{k},\boldsymbol \tau}\,e^{-iuH}\rho_{0^{-}}\mathcal{V}^{\dagger}_{\mathbf{k},\boldsymbol \tau}\right] \right\rangle}\,.
\end{equation}

Moreover, as usual in stochastic thermodynamics both for classical and quantum systems, the $\ell$-th statistical moment of the fluctuation quantity of interest (the heat $Q$ in our case) can be directly by means of the following formula:
\begin{equation}\label{work_moments}
\overline{\langle Q^{\ell}\rangle } = \left.(-i)^{\ell}\partial^{\ell}_{u}G(u)\right|_{u=0} \,,
\end{equation}
where $\partial^{\ell}_{u}(\cdot)$ denotes the $\ell$-th partial derivative of $(\cdot)$ with respect to $u$.

Finally, as already observed in Ref.\,\cite{GherardiniPRE2018}, the derivation of the heat characteristic function $G(u)$ shown above has the good property of being valid also if a protocol of Positive Operator-Valued Measures (POVMs), excluding the first and the last energy projective measurements, is applied to the quantum system. This because the full characterization of all the intermediate quantum measurements is encoded in the super-operator $\mathcal{V}_{\mathbf{k},\boldsymbol \tau}$ that is indeed present in the final expression of the heat characteristic function.

\subsection{Fluctuation theorem}

In the introduction we already mentioned that the Jarzynski identity takes also the acronym of \emph{fluctuation theorem}. The latter denotes the existing link among a non-equilibrium quantity, as the exchange energy, and the equilibrium free-energy difference. In this paragraph, we show that a quantum fluctuation theorem can be attained also when a quantum system is subject to a sequence of repeated projective measurements, and we prove that it holds independently on the presence of both intermediate measurements in the applied protocol and randomness in the waiting times $\boldsymbol \tau$ between the measurements. Consequently, this entails that the random distribution of the waiting times cannot be detected by evaluating the average value $\overline{\langle e^{-\beta Q} \rangle}$, with $\rho_{0^{-}}$ Gibbs thermal state, whatever are the values taken by $\boldsymbol \tau$ and $p(\boldsymbol \tau)$.

It is known \cite{Campisi2011PRE} that, for a quantum system prepared in the Gibbs thermal state
\begin{equation}
\rho_{0^{-}} = \frac{e^{-\beta H_0}}{{\rm Tr}[e^{-\beta H_0}]}\,,
\end{equation}
with $\beta$ inverse temperature and subject to a time-dependent forcing protocol in the time interval $[0,T]$, the Jarzynski equality is identically equal to
\begin{equation}\label{eq:standard:Jarzy}
\langle e^{-\beta (E'_m-E_n)}\rangle = e^{-\beta \Delta F}\,,
\end{equation}
independently on the number $M$ of intermediate projective measurements applied at regular times. In Eq.\,(\ref{eq:standard:Jarzy}), $E'_m$ denote the final eigenvalues of the time-dependent system Hamiltonian $H_t$ and
\begin{equation}
\Delta F \equiv - \beta^{-1}\ln \frac{{\rm Tr}[e^{-\beta H_{T}}]}{ {\rm Tr}[e^{-\beta H_{0}}]}
\end{equation}
is the free-energy difference. Accordingly, if the time-dependent forcing is turned off, then $\Delta F = 0$ such that
\begin{equation}\label{eq:av-exp(-Qq)}
\langle e^{-\beta Q}\rangle = 1\,,
\end{equation}
where we recall that $\langle e^{-\beta Q}\rangle$ denotes the quantum-mechanical expectation of $e^{-\beta Q}$ with fixed waiting times. Once again, note that in this specific case, corresponding to the one we are here discussing, all the (internal) energy variation $\Delta U$ in the quantum system has to be ascribed to heat.

Now, we are going to show that the fluctuation theorem (\ref{eq:av-exp(-Qq)}) remains valid also if the waiting times are sampled by the probability distribution $p(\boldsymbol \tau)$. To see this, one just need to evaluate the characteristic function (\ref{G_u_finale}) in $u=i\beta$ by taking the initial state $\rho_{0^{-}}$ as the Gibbs thermal state with inverse temperature $\beta$. Formally, it holds that
\begin{eqnarray*}
G(i\beta) &=& \overline{\langle e^{-\beta Q} \rangle} = \int d^{M}\boldsymbol \tau  p(\boldsymbol \tau)\sum_{\mathbf{k}}
{\rm Tr}\left[e^{-\beta H}\,\mathcal{V}_{\mathbf{k},\boldsymbol \tau}\,e^{\beta H}\,\frac{e^{-\beta H}}{Z}\mathcal{V}^{\dagger}_{\mathbf{k},\boldsymbol \tau}\right]\nonumber \\
&=& {\rm Tr}\left[\frac{e^{-\beta H}}{Z} \int d^{M}\boldsymbol \tau\,p(\boldsymbol \tau) \sum_{\mathbf{k}}\mathcal{V}_{\mathbf{k},\boldsymbol \tau}\mathcal{V}^{\dagger}_{\mathbf{k},\boldsymbol \tau}\right] = \frac{{\rm Tr}\left[e^{-\beta H}\right]}{Z} = 1\,.
\end{eqnarray*}
In fact, thanks to the cyclicity of the trace operation, the unitarity of the quantum evolutions between consecutive measurements, the idempotence of the measurement projectors, and the normalisation $\int d^{M}\boldsymbol \tau\,p(\boldsymbol \tau)=1$, the property
\begin{equation}\label{eq:property_q_maps}
\int d^{M}\boldsymbol \tau\,p(\boldsymbol \tau) \sum_{\mathbf{k}}\mathcal{V}_{\mathbf{k},\boldsymbol \tau}\mathcal{V}^{\dagger}_{\mathbf{k},\boldsymbol \tau} = \mathbb{I}
\end{equation}
is always valid for each possible value of the involved variables. In this regard, it is worth noting that Eq.\,(\ref{eq:property_q_maps}) is a direct consequence of the \emph{unitality} of the quantum map $\Phi$ that describes the evolution (from $t=0$ to $t=T$) of the system subject to a sequence of projective measurements randomly distributed over time.
It is indeed the unitality of $\Phi$ that ensures the validity of the fluctuation relation (\ref{G_u_finale}) that has recently found experimental confirmation in Ref.\,\cite{HernandezPRR2020}.

\subsection{Infinite-temperature thermalization}

In this paragraph, the asymptotic behaviour of the monitored quantum system is studied in the limit $M \gg 1$.

In doing this, let us consider the non-equilibrium protocol of $M$ intermediate projective measurements on the observable $\mathcal{O}$, and the probability $\widetilde{\pi}_{k_M}$ to detect the quantum system in the state $|\alpha_k\rangle\!\langle\alpha_k|$ after the $M$-th measurement. Thus, by denoting $\pi_{k_1}$ as the probability that the system has been projected in the projector $|\alpha_1\rangle\!\langle\alpha_1|$ by the $1$-th measurement on $\mathcal{O}$, $\widetilde{\pi}_{k_M}$ is returned by the formula
\begin{equation}
    \widetilde{\pi}_{k_M} = \sum_{k_1} \pi_{k_M|k_1} \pi_{k_1}\,,
\end{equation}
where
\begin{equation}\label{tildepi}
    \pi_{k_M|k_1} = \int d^{M} \boldsymbol\tau\,p(\boldsymbol{\tau}) \sum_{k_{1}, \dots, k_{M-1}} {\rm Tr}\left[ \mathcal{V}_{\mathbf{k},\boldsymbol\tau} |\alpha_{k_1}\rangle\!\langle\alpha_{k_1}| \mathcal{V}^{\dagger}_{\mathbf{k},\boldsymbol\tau}\right]
\end{equation}
denotes the conditional probability to measure the outcome $\alpha_{k_M}$ from the $M$-th measurement of $\mathcal{O}$ provided that the outcome from the first intermediate-measurement was $\alpha_{k_1}$. Let us observe that the expression of $\pi_{k_M|k_1}$ in Eq.\,(\ref{tildepi}) remains valid independently on the procedure adopted to possibly take into account the presence of initial coherence/correlation terms in $\rho_{0^{-}}$.

The first observation that allows to derive most of the results in Ref.\,\cite{GherardiniPRE2021} is that the expression of the conditional probability $\pi_{k_M|k_1}$ can be further simplified as
\begin{equation} \label{matrixnotation}
    \pi_{k_M|k_1} = \int d^{M} \boldsymbol\tau\,p(\boldsymbol{\tau}) \langle\alpha_{k_M}|\prod^M_{j=2} L(\tau_{j-1}) |\alpha_{k_1}\rangle\,,
\end{equation}
where the linear operator $L$ is implicitly defined by the following relation:
\begin{equation}\label{eq:relation_L_U}
    \langle\alpha_{k_{j-1}}| L(\tau_{j-1}) |\alpha_{k_j}\rangle \equiv \left| \langle\alpha_{k_{j-1}}| U(\tau_{j-1}) |\alpha_{k_j}\rangle \right|^{2}.
\end{equation}
Let us observe that $|\langle\alpha_{k_{j-1}}| U(\tau_{j-1}) |\alpha_{k_j}\rangle|^{2}$ is the $1$-th step conditional probability to measure the outcome $\alpha_{k_j}$ from the $j$-th projective measurement once measured the outcome $\alpha_{k_{j-1}}$ from the $(j-1)$-th one. From this consideration one can evince that, while the structure governing the occurrence of the outcomes from the intermediate measurements of $\mathcal{O}$ is Markovian by construction, the evolution of the quantum system's state can be in general non-Markovian, thus admitting a multi-step dependence on past states, usually called memory effects. This evidence has been recently shown in Ref.\,\cite{GherardiniArXiv2021_NM} thanks to the introduction of a stochastic generalization of the so-called transfer-tensor formalism \cite{CerrilloPRL2014,PollockQuantum2018}. Being $|\langle\alpha_{k_{j-1}}| U(\tau_{j-1}) |\alpha_{k_j}\rangle|^{2}$ a conditional probability, in matrix notation each $L(\tau)$ is the \emph{transition} matrix of a discrete-time Markov chain whose states are the eigenstates of the observable $\mathcal{O}$. This means that the matrices $L(\tau)$ are \emph{stochastic} and thus admit rows or columns summing to $1$. Moreover, since each element of $L(\tau_j)$ is provided by the square modulus of the corresponding element of a unitary matrix (always depending on $\tau_j$), the $L(\tau_j)$'s are also \emph{unistochastic}. This entails that all the eigenvalues $\lambda_k$ of each $L(\tau_j)$ are such that $|\lambda_k| \leq 1$ and at least one of them is equal to $1$. Therefore, as largely discussed in \cite{GherardiniPRE2021}, in the limit $M \gg 1$ the product of the transition matrices $L(\tau)$ tends asymptotically to the projector $\mathcal{P}_{\lambda=1}$ that spans the eigenspace identified by $\lambda=1$, i.e.,
\begin{equation}
L(\tau)^M \rightarrow \mathcal{P}_{\lambda=1}\,.
\end{equation}

Now, let us provide the explicit expressions of $\mathcal{P}_{\lambda=1}$. For this purpose, we look for the eigenvector $|v\rangle$ that satisfies the eigenvalue equation $L(\tau)|v\rangle = |v\rangle$ for all values of $\tau$, namely we look for the fixed point of $L(\tau)$. As starting point, let us assume that $\lambda=1$ is non degenerate with the result that $L(\tau)^{M-1} \rightarrow |v\rangle\!\langle v|$. Under this hypothesis, thanks to the symmetry of $L(\tau)$, the eigenvalue equation $L(\tau)|v\rangle = |v\rangle$ is uniquely solved by
\begin{equation}
|v\rangle = \frac{1}{\sqrt{N}}\sum_{k=1}^{N}\,|\alpha_k\rangle
\end{equation}
that, quite remarkably, does not depend on $\tau$. Thus, the stochastic process (underlying the dynamics of the quantum system subject to randomly distributed quantum measurements) is \emph{ergodic}, and admits the unique asymptotic configuration whereby the probabilities that the final state of the system is one of the eigenvectors $|\alpha_k\rangle$ of $\mathcal{O}$ are all the same.

The consequence of this on the conditional probability $\pi_{k_M|k_1}$ is that, in the $M \gg 1$ limit, the conditional probability $\pi_{k_M|k_1}$ simplifies as
\begin{equation}
    \pi_{k_M|k_1} = \langle\alpha_{k_M}|v\rangle \langle v|\alpha_{k_1}\rangle = \frac{1}{N}
\end{equation}
such that
\begin{equation}
    \widetilde{\pi}_{k_M} = \sum_{k_1}\pi_{k_M|k_1}\pi_{k_1} = \frac{1}{N}\,,
\end{equation}
irrespective of the state onto which the quantum system collapses after the first measurement of $\mathcal{O}$, meaning that for $M$ large the information on the initial condition is lost. As a result, for increasing values of $M$, the system's state after the last measurement on $\mathcal{O}$ is $\rho_M = \mathbb{I}/N$, i.e., the maximally mixed state. This also implies that, being $\rho_M$ diagonal in every basis by definition of the maximally mixed state, the $2$-nd energy energy measurement required by the TPM scheme does yield no effects and also the final energy outcomes are equiprobable. In \cite{GherardiniPRE2021}, such asymptotic behaviour of a generic quantum system subject to repeated projective measurements has been interpreted as an effective thermalization process towards a thermal state with infinite temperature $T=\infty$ ($\beta=0$), where the measurement apparatus, which probes the system in a stroboscopic manner, acts as a thermal reservoir with energy ideally tending to infinite.

\subsection{Partial thermalization \& quantum Zeno effect}

Exceptions to the infinite-temperature thermalization, leading to the so-called \emph{partial thermalization} \cite{GherardiniPRE2021}, occur due to a degeneracy of the eigenvalue $\lambda=1$ of the transition matrix $L(\tau)$. Specifically, we briefly discuss two main exceptions: (i) $\mathcal{O}$ and $H$ share a common invariant subspace; (ii) fluctuations originated in the quantum Zeno regime.

The first exception concerning partial thermalization can be experienced by removing the hypothesis that the greatest eigenvalue $\lambda=1$ of $L(\tau)$ is non-degenerate (for any value of $\tau$). Before that, let us consider the Perron-Frobenius theorem \cite{Perron} stating that, if the generic matrix $\mathcal{A}$ is both primitive (i.e., it is non-negative [all elements $\geq 0$] and its $k$-th powers are positive for some natural number $k$) and irreducible (namely, it cannot be put in a block diagonal form with a change of basis), then the greatest eigenvalue (in modulus) is real, positive and non-degenerate. In the case we are analysing, the transition matrix $L(\tau)$ is primitive; therefore, if $L(\tau)$ is irreducible, then the Perron-Frobenius theorem guarantees that its eigenvalue $\lambda=1$ is non-degenerate. This being said, let us assume that the measurement observable $\mathcal{O}$ and the Hamiltonian $H$ share a common non-trivial invariant subspace. As a consequence, in the basis $\{|\alpha_k\rangle\}_{k=1}^{N}$ defining the eigenstates of $\mathcal{O}$, $H$ takes the form of a block diagonal matrix, i.e.,
\begin{equation}
H =  \begin{pmatrix} H_1 & &  \\ & \ddots & \\ & & H_R \end{pmatrix}\,,
\end{equation}
with $R$ the number of blocks and $H_r$ irreducible Hermitian matrices acting on the subspaces $S_r$ with $r=1,\dots,R$. Accordingly, also the transition matrices $L(\tau_j)$ are block diagonal that take the form
\begin{equation}\label{eq:L_partial_ITT}
L(\tau_j) =  \begin{pmatrix} L_1 (\tau_j) & &  \\ & \ddots & \\ & &  L_R (\tau_j) \end{pmatrix}\,,
\end{equation}
where $L_r(\tau_j)$ are unistochastic irreducible operators acting on the subspaces $S_r$ for $r=1,\ldots,R$ and $j=1,\ldots,M$. Then, by applying the Perron-Frobenius theorem, one gets that no further degeneracy is present in each matrix $L_r(\tau_j)$ (being them irreducible matrices for any $r,j$), and to each subspace $S_r$ can associate the following set of eigenvectors:
\begin{equation}
|v_r\rangle \equiv \frac{1}{\sqrt{\dim{S_r}}} \sum_{k:|\alpha_k\rangle \in S_r} |\alpha_k\rangle\,.
\end{equation}
As a result, the system's state $\rho_{M}$ after the last (the $M$-th) measurement on $\mathcal{O}$ is no longer equal to the maximally mixed state $\mathbb{I}/N$ such that $\pi_{k_M|k_1}=1/\dim{S_r}$ if both $|\alpha_{k_1}\rangle$ and $|\alpha_{k_M}\rangle$ belongs to the same subspace $S_r$, and $\pi_{k_M|k_1}=0$ otherwise. Hence, in the former case, the probability $\widetilde{\pi}_{k_M}$ (to detect the system in $|\alpha_k\rangle\!\langle\alpha_k|$ after the $M$-th measurements on $\mathcal{O}$) keeps memory of the initial state. Moreover, one also determines that the eigenspaces associated to the eigenvalues of $L(\tau_j)$ are not mixed by the initial and final energy projective measurements of the TPM scheme. This is reflected in writing the probability $\widetilde{p}_m$ to measure the $m$-th energy value of $H$ at the end of the monitoring dynamics as
\begin{equation}
    \widetilde{p}_m = \frac{1}{\dim{S_r}}\sum_{n:|E_n\rangle \in S_r}p_n
\end{equation}
that corresponds to the final energy eigenvector $|E_m\rangle\in S_r$. In addition, regarding the heat characteristic function $G(u)$, it turns out that in the $M$-large limit it is provided by the sum of the corresponding heat characteristic functions in each subspace $S_r$, i.e.,
\begin{equation}
G(u) = \sum_{r=1}^{R} \frac{1}{\dim{S_r}}{\rm Tr}\left[\rho_{0^{-}}\,e^{-iH_{r}u}\right]{\rm Tr}\left[\rho_{M}\,e^{iH_{r}u}\right]
\end{equation}
such that in the limiting case of $R=N$ the standard result $G(u)=1$ is recovered. Therefore, all this discussion leads us to conclude that in the limit of $M \rightarrow \infty$, if the eigenvalue $\lambda=1$ of $L(\tau)$ is degenerate, there occur the complete mixing of only the eigenstates $|v_r\rangle$ that define and decompose the subspaces $S_{r}$, thus preventing infinite-temperature thermalization.

The second exception, we are going to discuss, concerns the case in which the value of the waiting times $\tau_j$ ($j=1,\ldots,M$) is on average much smaller than the inverse of the energy scale of the system provided, roughly speaking, by the largest eigenvalue (energy) of its Hamiltonian $H$ \cite{Gherardini2016NJP,Mueller2017ADP,KofmanNature2000,FacchiPRL2002,FacchiJPA2008,SmerziPRL2012,SchaferNatComm2014,SignolesNatPhys2014}. In such limiting case, the quantum Zeno effect is recovered and any thermalization process (both total and partial) is prevented. For simplicity, without loss of generality, we take non random waiting times with a constant value $\tau$ (thus, a regular sequence of projective measurements).

As shown in Ref.\,\cite{GherardiniPRE2021}, in the quantum Zeno regime the unitary operator $U(\tau)$ and the transition matrix $L(\tau)$ are nearly close to the identity matrix. To illustrate this, let us consider the $kj$-th elements of such operators. Regarding $U(\tau)$, in the Zeno regime we can find that
\begin{equation}\label{eq:Zeno_1}
\langle\alpha_k|U(\tau)|\alpha_{\ell}\rangle \approx \delta_{k,\ell} - i\tau\langle\alpha_k|H|\alpha_\ell\rangle + O(\tau^2)\,,
\end{equation}
where $\delta_{k,\ell}$ denotes the Kronecker delta ($\delta_{k,\ell}=1$ if $k=\ell$; $0$ otherwise), while the $kj$-th elements of $L(\tau)$ is simply provided by
\begin{equation}\label{eq:Zeno_2}
\langle\alpha_k| L(\tau) |\alpha_\ell\rangle \approx \delta_{k,\ell} + O(\tau^2)
\end{equation}
that one can justify by recalling the relation between $L(\tau)$ and $U(\tau)$ as given by Eq.\,(\ref{eq:relation_L_U}). In this way, since $\tau$ is constant as well as the total duration $M\tau$ of the monitoring dynamics, also the conditional probability $\pi_{k_M|k_1}$ of Eq.\,(\ref{matrixnotation}) can be developed in the limiting case of $\tau$ much smaller than the system's energy scale. Formally, given that $O(\tau^2) = O(M^{-2})$ (in the limit $M \gg 1$, indeed, the value of all the waiting times approaches to zero as $M^{-1}$), one gets
\begin{equation}
    \pi_{k_M|k_1} \approx \delta_{k_{1},k_{M}} + (M-1)O(M^{-2}) = \delta_{k_{1},k_{M}} + O(M^{-1})\,,
\end{equation}
thus proving that, in the quantum Zeno regime, a quantum system subject to many measurements of the observable $\mathcal{O}$ remains indefinitely frozen in one of the eigenstates of the measurement observable.

\section{Case studies}

Here, to illustrate a part of the theoretical results shown above, we apply them respectively to two- and three-quantum systems. Explicit calculations and numerical simulations are thus provided for each of these cases.

\subsection{Two-level quantum systems}

In this paragraph, we present the explicit derivation of the heat characteristic function $G(u)$ for a two-level quantum system, in accordance with the results in Ref.\,\cite{GherardiniPRE2018}. In this way, one can directly observe the dependence from the various parameters of the system on the way the heat statistics change, and how such parameters can be tuned to work out interesting thermodynamic quantities.

For this purpose, we consider the waiting times between the $M$ measurements, enclosed in the vector $\boldsymbol\tau$, to be independent and identically distributed (i.i.d.) random variables sampled from the probability distribution $p(\boldsymbol\tau)$. This means that the joint distribution of the waiting times is $p(\boldsymbol\tau) = \prod_{j=1}^{M}p(\tau_{j})$. We also assume each $p(\tau_j)$ to be a bimodal probability density function, with values $\tau^{(1)}$, $\tau^{(2)}$ and probabilities $p_{1}$ and $p_{2} = 1 - p_{1}$ for each $j=1,\ldots,M$. Moreover, let $E_{+}$ and $E_{-}$ denote the two energy eigenvalues of our system's Hamiltonian. Then, since energy fluctuations are evaluated by means of the TPM scheme, the initial density operator $\rho_{0^{-}}$ is taken as a diagonal matrix in the energy eigenbasis:
\begin{equation}\label{eq:rho_0_2_level_system}
\rho_{0^{-}} = c_{1}|E_{+}\rangle\!\langle E_{+}| + c_{2}|E_{-}\rangle\!\langle E_{-}|\,,
\end{equation}
with $c_1$, $c_2\in[0,1]$ and $c_2 = 1 - c_1$. Finally, as above, the eigenstates of the intermediate measurement observable $\mathcal{O}$ are denoted as $\{|\alpha_{k}\rangle\}$ with $k=1,2$. Without loss of generality, they can be expressed as a linear combination of the energy eigenstates, i.e.,
\begin{equation}
\begin{split}
&\ket{\alpha_1}=a\ket{E_{+}}-b\ket{E_{-}}\\
&\ket{\alpha_2}=b\ket{E_{+}}+a\ket{E_{-}}\,,
\label{eq:abasis}
\end{split}
\end{equation}
where $a,b\in\mathbb{C}$, $|a|^2+|b|^2=1$ and $a^{\ast}b = ab^{\ast}$.

Under these assumptions and by explicitly calculating the trace, we can rewrite the heat characteristic function $G(u)$ as
\begin{equation}\label{eq:B12}
G(u) = \sum_{j = 0}^{M-1}\binom{M-1}{j}f(u)\mathcal{L}(\tau^{(1)})^{j}\mathcal{L}(\tau^{(2)})^{M-j-1}g(u)\,p_1^{j}\,p_2^{M-j-1}
\end{equation}
and use Eq.\,(\ref{eq:B12}) to derive the explicit dependence of $G(u)$ from the various parameters $a$, $b$, $c_1$, $c_2$, $p_1$, $p_2$, $\tau_1$, $\tau_2$ and $E$ of the system dynamics. To this end, the system's energy values $E_{\pm}$ are taken equal to $\pm E$ and by making use of the energy eigenvalue equation, i.e., $H|E_{\pm}\rangle = E_{\pm}|E_{\pm}\rangle$, we get
\begin{equation}\label{B2}
f(u)^{T} = \begin{pmatrix}
\langle\alpha_{1}|e^{iuH}|\alpha_{1}\rangle \\
\langle\alpha_{2}|e^{iuH}|\alpha_{2}\rangle
\end{pmatrix} =
\begin{pmatrix}
|a|^{2}e^{iuE} + |b|^{2}e^{-iuE} \\
|a|^{2}e^{-iuE} + |b|^{2}e^{iuE}
\end{pmatrix},
\end{equation}
where we recall the elements $\{|\alpha_{k}\rangle\}$, $k = 1,2$, of the intermediate measurement basis are chosen as linear combinations of the energy eigenstates $|E_{\pm}\rangle$. Another quantity that we need to explicitly express is the transition matrix $\mathcal{L}$  that turns out to be
\begin{equation}\label{B3}
\mathcal{L} = \begin{pmatrix}
\left||a|^{2}e^{-iEt} + |b|^{2}e^{iEt} \right|^{2} &
\left|a^{\ast}b\,e^{-iEt} - ab^{\ast}e^{iEt} \right|^{2} \\ \left|b^{\ast}a\,e^{-iEt} - ba^{\ast}e^{iEt} \right|^{2} & \left||b|^{2}e^{-iEt} + |a|^{2}e^{iEt} \right|^{2}
\end{pmatrix}\nonumber = \begin{pmatrix}
1-\overline{\nu} & \overline{\nu} \\ \overline{\nu} & 1-\overline{\nu}
\end{pmatrix},
\end{equation}
where
\begin{equation}\label{eq:nu}
\overline{\nu} \equiv 2|a|^{2}|b|^{2}\sin^{2}(\langle\tau\rangle\delta E)\,,
\end{equation}
$\delta E \equiv(E_{+} - E_{-}) = 2E$ and $\langle\tau\rangle \equiv \sum_{k}\tau^{(k)}p_{k}$. Then, by resorting to the decomposition (\ref{eq:rho_0_2_level_system}) of the initial density operator $\rho_{0^{-}}$ in the energy basis of the system and again Eq.\,(\ref{eq:abasis}), one has that
\begin{equation}\label{eq:g(u)}
g(u) = \begin{pmatrix}
\langle\alpha_{1}|e^{-iuH}\rho_{0^{-}}|\alpha_{1}\rangle \\
\langle\alpha_{2}|e^{-iuH}\rho_{0^{-}}|\alpha_{2}\rangle
\end{pmatrix}
= \begin{pmatrix}
|a|^{2}c_{1}e^{-iuE}+|b|^{2}c_{2}\,e^{iuE}\\
|a|^{2}c_{2}\,e^{iuE} + |b|^{2}c_{1}e^{-iuE}
\end{pmatrix}.
\end{equation}

In conclusion, if we substitute in Eq.\,(\ref{eq:B12}) the expressions of $f(u)$, $\mathcal{L}$ and $g(u)$ as given by Eqs.\,(\ref{B2}), (\ref{B3}), (\ref{eq:nu}) and (\ref{eq:g(u)}), the explicit dependence of $G(u)$ from the parameters of the system is given by the following equation:
\begin{eqnarray}\label{eq:G_complete_expr}
G(u)&=&\sum_{j = 0}^{M-1}\binom{M-1}{j}\begin{pmatrix}
|a|^2e^{iuE}+|b|^2e^{-iuE}\\
|a|^2e^{-iuE}+|b|^2e^{iuE}
\end{pmatrix}^{T}\begin{pmatrix}
1-\nu_{1} & \nu_{1} \\ \nu_{1} & 1-\nu_{1}
\end{pmatrix}^{j} \begin{pmatrix}
1-\nu_{2} & \nu_{2} \\ \nu_{2} & 1-\nu_{2}
\end{pmatrix}^{M-j-1} \nonumber \\
&\cdot& \begin{pmatrix}
|a|^2c_{1}e^{-iuE}+|b|^2c_{2}\,e^{iuE}\\
|a|^2c_{2}\,e^{iuE}+|b|^2c_{1}e^{-iuE}
\end{pmatrix} p_1^{j}\,p_2^{M-j-1}\,,
\end{eqnarray}
where $\nu_{k} \equiv \nu(\tau^{(k)}) = 2|a|^{2}|b|^{2}\sin^{2}(2\tau^{(k)}E)$.

Looking at the expression in Eq.\,(\ref{eq:G_complete_expr}), in Ref.\,\cite{GherardiniPRE2018} it has been proven that $G(u)$ has a \emph{discontinuity} under the limits of $|a|^2\rightarrow 0,1$ and $M\rightarrow\infty$ that are not commuting. In fact, when $|a|^2\rightarrow 0,1$ and a finite number $M$ of intermediate measurements on $\mathcal{O}$ is performed, $G(u)$ is identically equal to $1$; while for $M\rightarrow\infty$ the characteristic function does not longer depend on $a$ and its expression becomes
\begin{equation}\label{G_u_qu}
G(u) = \frac{(1+e^{2iuE})}{2} - c_{1}\sinh(2iuE)\,.
\end{equation}
To obtain this, let us take Eq.\,(\ref{eq:B12}) with $a \neq 0$ and, then, use the binomial theorem given by the well-known relation
\begin{equation}
\displaystyle{(x+y)^{n} = \sum_{\ell=0}^{n} \binom{n}{\ell}x^{n-\ell}y^{\ell}}\,,
\end{equation}
with $x$, $y$ arbitrary real variables. As a result,
\begin{equation}
G(u) = f(u)\left(p_{1}\,\mathcal{L}(\tau^{(1)}) + p_{2}\,\mathcal{L}(\tau^{(2)})\right)^{M-1}g(u)\,.
\end{equation}
Now, if we introduce the quantity $\zeta\equiv p_{1}\nu_1+p_{2}\nu_2$, then the weighted sum with respect to $p(\tau)$) of $\mathcal{L}(\tau^{(1)})$ and $\mathcal{L}(\tau^{(2)})$ -- to be seen effectively as transition matrices -- can be simplified as
\begin{equation}\label{eq:weighted_sum_trans_matrices}
p_{1}\,\mathcal{L}(\tau^{(1)}) + p_{2}\,\mathcal{L}(\tau^{(2)}) =
\begin{pmatrix}
1-\zeta & \zeta \\ \zeta & 1-\zeta
\end{pmatrix},
\end{equation}
whose eigenvalues are $1$ and $(1-2\,\zeta)\leq 1$. Hence, in the limit $M\rightarrow\infty$, the weighted sum (\ref{eq:weighted_sum_trans_matrices}) of the transition matrices $\mathcal{L}(\tau^{(1)})$ and $\mathcal{L}(\tau^{(2)})$ tends to a projector and $G(u)$ is provided effectively by Eq.\,(\ref{G_u_qu}).

Finally, having analytically derived the heat characteristic function $G(u)$ for a generic two-level quantum system subject to repeated quantum measurements (randomly distributed), we can quite easily compute also its (partial) derivatives $\partial^{n}_{u}G(u)$ with respect to $u$. In this way, one is able to derive all the statistical moments of the heat probability distribution. As a result of this calculation, the $n-$th order derivative of the heat characteristic function equals to
\begin{equation}\label{eq:partial_derivative_annealed}
\partial^{n}_{u}G(u) = \displaystyle{\sum_{j = 0}^{M-1}\sum_{\ell=0}^{n}A^{\ell}(u)^{T}}
\begin{pmatrix}
1-\nu_1 & \nu_1 \\
\nu_1 & 1-\nu_1
\end{pmatrix}^{j}
\begin{pmatrix}
1-\nu_2 & \nu_2 \\
\nu_2 & 1-\nu_2
\end{pmatrix}^{M-j-1} B^{n-\ell}(u)\,p_{1}^{j}\,p_{2}^{M-j-1}\,,
\end{equation}
where
\begin{equation}
A^{\ell}(u) \equiv
(i)^{\ell}
\begin{pmatrix}
\bra{\alpha_{1}}H^{\ell}e^{iuH}
\ket{\alpha_{1}} \\
\bra{\alpha_{2}}H^{\ell}e^{iuH}
\ket{\alpha_{2}}
\end{pmatrix}
\end{equation}
and
\begin{equation}
B^{\ell}(u)\equiv
(-i)^{\ell}
\begin{pmatrix}
\bra{\alpha_{1}}H^{\ell}e^{-iuH}\rho_{0^{-}}
\ket{\alpha_{1}} \\
\bra{\alpha_{2}}H^{\ell}e^{-iuH}\rho_{0^{-}}
\ket{\alpha_{2}}
\end{pmatrix}.
\end{equation}
That concludes the example on two-level quantum systems that, albeit being quite simple in nature, do a good job of illustrating the theoretical findings and showing how to concretely apply them for the statistical and thermodynamic study of physical systems.

\subsection{Three-level quantum systems}

To introduce a paragraph on $3$-level quantum systems (i.e., $N=3$, with $N$ denoting the number of system's levels), first let us explain why taking just a $2$-level system may be insufficient to provide general results. In this regard, it is worth noting that, while for any $2$-level quantum system in a mixed state is always possible to formally define an inverse temperature $\beta$ \cite{footnote1}, this is no longer true for $N>2$ since the number of parameters needed to specify a mixed state is in general $N-1$. However, in Ref.\,\cite{GiachettiCM2020} it has been shown that an unique, non-zero value $\beta_{\rm eff}$ exists such that the condition $G(i \beta_{\rm eff})=1$ is fulfilled. From the point of view of energy fluctuations, $\beta_{\rm eff}$ can be interpreted as an effective inverse temperature associated to the initial state $\rho_{0^{-}}$. Below, we specifically focus on the case $N=3$, by showing how $\beta_{\rm eff}$ can be estimated.

As in the previous paragraph concerning $2$-level quantum systems, the initial density operator $\rho_{0^{-}}$ is taken as a diagonal matrix in the energy eigenbasis, i.e., $\rho_{0^{-}} = \sum_{k}c_{k}|E_k\rangle\!\langle E_{k}|$. Then, let us start proposing a suitable parametrization of the initial probabilities $c_1, c_2, c_3$. While for a $2$-level system the ratio $\frac{c_2}{c_1} = e^{-\beta_{\rm eff}(E_2 - E_1)}$ uniquely defines the system's temperature, in the $N=3$ case we have to introduce the three parameters $b_1,b_2,b_3$ to express the ratios $\frac{c_2}{c_1}$, $\frac{c_3}{c_2}$ and $\frac{c_1}{c_3}$, respectively:
\begin{equation}
\frac{c_2}{c_1} = e^{- b_1 (E_2-E_1)},\hspace{1cm}\frac{c_3}{c_2} = e^{- b_2 (E_3-E_2)}, \hspace{1cm} \frac{c_1}{c_3} = e^{- b_3 (E_1-E_3)}
\end{equation}
such that $b_k = \beta$, $\forall k=1,2,3$, for the thermal state $e^{-\beta H}/Z$. Notice that the parameters $b_1, b_2, b_3$ are not independent, since the product of the above ratios is fixed to $1$. In fact, the $b_k$'s are constrained by the following relation:
\begin{equation}\label{orthogonality}
\sum^3_{k=1} b_k \Delta_k = 0 \ ,
\end{equation}
where $\Delta_1 \equiv E_2 - E_1$, $\Delta_2 \equiv E_3 - E_2$ and $\Delta_3 \equiv E_1 - E_3$. Let us observe that, by definition, $\sum^3_{k=1} \Delta_k = 0$. This means that, when the initial state is thermal and the vector $(b_1,b_2,b_3)$ is proportional to $(1,1,1)$, the condition \eqref{orthogonality} is automatically satisfied. Accordingly, we expect in general that one can conveniently parametrize $b_k$ in terms of the \emph{orthogonal} and \emph{parallel} components to the vector $(1,1,1)$, which can be interpreted as the thermal and non-thermal components of the initial state, respectively. Formally,
\begin{equation}
(b_1, b_2, b_3) = \beta (1,1,1) + \frac{\alpha}{v} (\Delta_3 - \Delta_2,\,\Delta_1 - \Delta_3,\,\Delta_2 - \Delta_1)\,,
\end{equation}
with $\alpha$ deniting the distance to a thermal state and $v$ acting as a normalization constant:
\begin{equation}
v^2 = 3\left
(\Delta_1^2 + \Delta_2^2 + \Delta_3^2\right).
\end{equation}
In this way, the coefficients $c_k$'s are given by
\begin{equation}\label{eq:coeffs_c}
c_1 = \frac{1}{\Tilde{Z}} \exp\left[-\beta E_1 + \frac{\alpha}{v} (E_2 - E_3)^2\right], \hspace{0.4cm} c_2 = \frac{1}{\Tilde{Z}} \exp\left[-\beta E_2 + \frac{\alpha}{v}(E_3 - E_1)^2\right], \hspace{0.4cm} c_3 =\frac{1}{\Tilde{Z}} \exp\left[- \beta E_3 + \frac{\alpha}{v}(E_1 - E_2)^2\right],
\end{equation}
where
\begin{equation}\label{pseudoZ}
\Tilde{Z} = \Tilde{Z} (\alpha, \beta) \equiv e^{-\beta E_1 + \frac{\alpha}{v} (E_2 - E_3)^2} +  e^{- \beta E_2 + \frac{\alpha}{v}(E_3 - E_1)^2} + e^{- \beta E_3 + \frac{\alpha}{v}(E_1 - E_2)^2}
\end{equation}
is a \emph{pseudo-partition function} ensuring the normalization of the initial density operator \cite{GiachettiCM2020}.

Now, without loss of generality, we choose the zero of the energy such that $E_2=0$, with $E_3>0$ and $E_1 <0$. Moreover, we reduce our analysis to the $\beta > 0$ region, since our choice of parameters for the parametrization of $\rho_{0^{-}}$ is left unchanged by the transformation $\{\beta \rightarrow - \beta,\,E_k \rightarrow - E_k\}$. Hence, the case of $\beta < 0$ is equivalent to the $\beta >0$ case in the fictitious system with $E^{\prime}_k = -  E_k$, which also leaves unchanged the condition $E_2=0$.
As discussed above, in the limit of large $M$ (and $\tau_j$'s finite) the quantum system ends up in the maximally mixed state (thermal state with $\beta=0$), with the exception of the case in which some of the eigenstates of the Hamiltonian $H$ and the measurement observable $\mathcal{O}$ coincide. This allows us to derive an analytic expression of $G(i\epsilon)$, heat characteristic function after the substitution $u=i\epsilon$. Here, the real variable $\epsilon$ has to meant just as a scale factor making $G(i\epsilon)$ adimensional. Thus, by considering in the large-$M$ limit the final state of the $3$-level quantum system independent from the initial state, one has that joint probability $p_{m,n} \equiv p_{m|n}p_{n}$ to measure the $n$-th initial and $m$-th energy outcomes by means of the TPM scheme is simply equal to $p_{m,n} = c_{n}/3$. Hence,
\begin{equation} \label{G}
G(i\epsilon) = \me{e^{- \epsilon\,Q}} = \frac{1}{3}\sum_{m=1}^3 e^{- \epsilon\,E_m} \sum_{n=1}^3 c_{n}\,e^{ \epsilon\,E_n} \,,
\end{equation}
which in turn can be expressed in terms of the actual partition function $Z(\beta) \equiv {\rm Tr}[e^{-\beta H}]$ and the pseudo-partition function introduced in Eq.\,$\eqref{pseudoZ}$:
\begin{equation} \label{GZ}
G(i\epsilon;\alpha,\beta) = \frac{Z(\epsilon)}{Z(0)} \frac{\Tilde{Z}(\alpha, \beta - \epsilon)}{\Tilde{Z}(\alpha,\beta)} \ .
\end{equation}
Regardless of the choice of the system's parameters, for $\alpha=0$ (initial thermal state at inverse temperature $\beta$) we recover the known fluctuation relations $G(0)=1$ and $G(i\beta)=1$. Moreover, we can also notice that an analytical expression for the non-trivial solution $\beta_{\rm eff}$ obeying $G(i\beta_{\rm eff}) = 1$ can be obtained only numerically.

\begin{figure}[h!]
    \centering
    \includegraphics[scale=0.575]{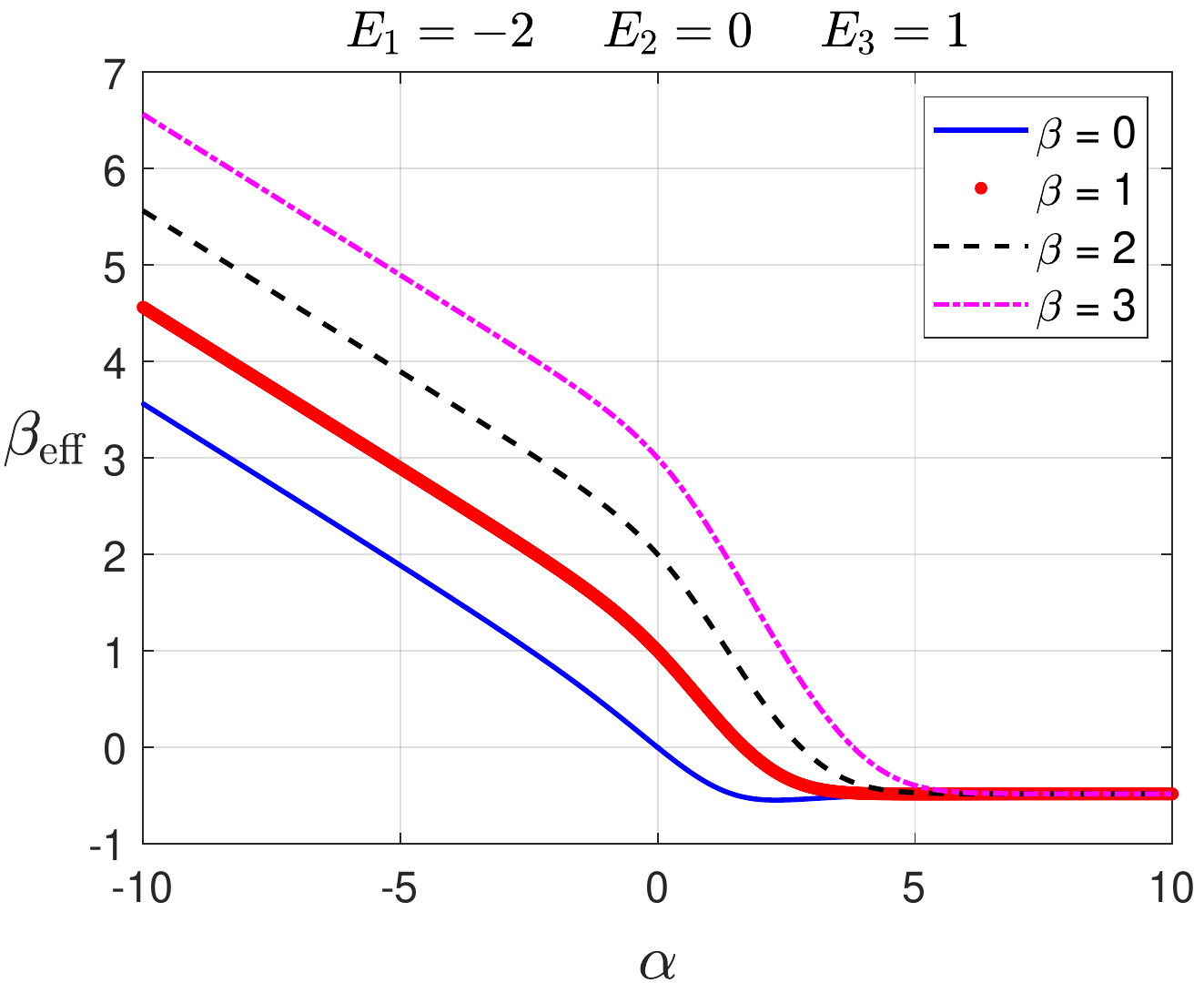}
    \caption{Numerical estimates of $\beta_{\rm eff}$ as a function of $\alpha$, with $\beta \in \{0, 1, 2, 3\}$ and energy values $\{E_1 = -2,\,E_2 = 0,\,E_3 = 1\}$.}
    \label{fig:betaeff}
\end{figure}

Thus, in Fig.\,\ref{fig:betaeff} we numerically compute $\beta_{\rm eff}$ as a function of $\alpha$ (i.e., the non-thermal component of $\rho_{0^{-}}$) for different values of $\beta > 0$. The following energy values are chosen: $\{E_1 = -2,\,E_2 = 0,\,E_3 = 1\}$ that represents the case with $E_3 - E_2 < E_2 - E_1$. Then, the energy unit is chosen such that the smallest energy gap is set to $1$. As expected, for $\alpha = 0$ we get $\beta_{\rm eff} = \beta$, independently on the values taken by the energies $\{E_1,E_2,E_3\}$. Moreover, the asymptotic value $\overline{\beta}_{\rm eff}$ of $\beta_{\rm eff}$ for large positive values of $\alpha$ (corresponding of having as initial density operator the pure state $\rho_{0^{-}} = |E_2\rangle\!\langle E_2|$) is provided by the only solution of the equation
\begin{equation} \label{asymptotic}
e^{\overline{\beta}_{\rm eff} (E_2 - E_1)} + e^{-\overline{\beta}_{\rm eff} (E_3-E_2)} = 2
\end{equation}
such that one gets $\overline{\beta}_{\rm eff} = 0$ if $E_1 = - E_3$. In this way, we can also derive some limiting values for $\bar{\beta}_{\rm eff}$, i.e.,
\begin{equation} \label{limits}
- \frac{\ln{2}}{E_3 - E_2} < \overline{\beta}_{\rm eff} < \frac{\ln{2}}{E_2-E_1}\,,
\end{equation}
which are met respectively in the limits $E_3 - E_2 \gg E_2 - E_1$ and $E_3 - E_2 \ll E_2 - E_1$.
Finally, the divergence for $\alpha \rightarrow - \infty$ in Fig.\,\ref{fig:betaeff} can be understood by noticing that in this limit (with $\beta \neq 0$) $\rho_{0^{-}} \asymp |E_1\rangle\!\langle E_1|$ when $E_3 - E_2 < E_2 - E_1$ or $\rho_{0^{-}} \asymp |E_3\rangle\!\langle E_3|$ if $E_3 - E_2 > E_2 - E_1$, which one can interpret as thermal state with zero temperature. As extensively shown in Ref.\,\cite{GiachettiCM2020}, for large negative values of $\alpha$ the asymptotic behavior of $\beta_{\rm eff}$  is linear with $\alpha$, i.e., $\beta_{\rm eff} \approx r \alpha$ with $r$ simply provided by the following relation:
\begin{equation}
r = \frac{E_1 + E_3 - 2 E_2}{v}
\end{equation}
that just depends on the energy values and the normalization constant $v$.

\section{Conclusions}

In this paper we have presented a general discussion on the characterization of the statistics of energy fluctuations in monitored quantum systems subject to quantum measurements. We considered quantum systems subject to a sequence of projective measurements performed both at regular and random time instants. Since the quantum system Hamiltonian is taken as a time-independent operator, energy variations originated by monitoring the system are ascribed as heat exchanged with the external measurement apparatus. In this regard, it is worth noting that we have not explicitly taken into account the energetic cost in performing a projective measurement \cite{DeffnerPRE2016,AbdelkhalekArXiv2018} onto the eigenstates of a generic observable, a topic we plan to address in the future.

Here, for the analysed open quantum system dynamics, we have showed the validity of the quantum fluctuation relation $\langle e^{-\beta Q}\rangle=1$ with $\beta$ inverse temperature of an initial Gibbs thermal state. The fluctuation relation is valid independently on the presence of both intermediate measurements during the dynamics and randomness in the waiting time between measurements. This result finds justification in the fact that overall the dynamics is well-described by a unital quantum map obeying Eq.\,(\ref{eq:property_q_maps}). It is worth noting that most of the presented findings are based on the possibility to recover the analytical expression of the heat characteristic function $G(u)$. This powerful mathematical tool, indeed, allows to access all the heat statistics, by just computing derivatives of $G(u)$ and evaluating them at specific points of its domain. Furthermore, under the previous assumptions (i.e., initial Gibbs states and unital quantum maps), one can determine analytically also the asymptotic behaviour of the monitored quantum system, in the limits of a large number of intermediate quantum measurements and an increasing number of system's levels. While the latter is discussed in \cite{GherardiniPRE2021}, for the former we here explain that the asymptotic state is the maximally mixed one in case the measurement observable $\mathcal{O}$ and the system's Hamiltonian $H$ share a common invariant subspace. Otherwise, thermalization processes at lower temperatures occur. In these regimes, we expect to observe non-trivial quantum effects in subspaces within the whole system Hilbert space, which one could likely observe also in commercial quantum circuits as IBM or Rigetti \cite{SolfanelliPRXQ2021}.

As future perspective of the addressed topics, it will surely deserve consideration to extend the presented theory to non-unital dynamics. A recent effort in this direction has been conducted in Ref.\,\cite{HernandezArXiv2021} where a unified framework to describe energy fluctuations in quantum systems under controllable feedback mechanisms is presented. In such a case, indeed, the considered experimental system (a NV center in diamond) is repeatedly subject to random measurements and dissipative events (modelled by superoperator in Lindblad form \cite{PetruccioneBook}), whose occurrence is conditioned by the measurements' results. Moreover, the interplay among quantum measurements and feedback control, giving rise to non-unital quantum dynamics, is responsible for asymptotic non-equilibrium steady-states with non-zero stationary coherence terms, with strong similarities with the ones derived in Ref.\,\cite{GuarnieriPRL2018}.

\subsection*{Acknowledgments}

Discussions with A. Belenchia, M. Campisi, P. Cappellaro, F.S. Cataliotti, D. Cohen, N. Fabbri, S. Hern\'andez-G\'omez, M. Paternostro, F. Poggiali and A. Sone are gratefully acknowledged. The authors acknowledge the MISTI Global Seed Funds MIT-FVG collaboration grants ``NV centers for the test of the Quantum Jarzynski Equality (NVQJE)'' and ``Non-Equilibrium Thermodynamics of Dissipative Quantum Systems'', and the MIUR-PRIN2017 project ``Coarse-grained description for non-equilibrium systems and transport phenomena (CO-NEST)'' No. 201798CZL. S.G. also acknowledges The Blanceflor Foundation for financial support through the project ``The theRmodynamics behInd thE meaSuremenT postulate of quantum mEchanics (TRIESTE)'' awarded in 2021. Support form the Progetto Bilaterale CNR/RS ``Testing fundamental theories with ultracold atoms'' is also acknowledged.


\begin{thebibliography}{99}
\bibitem{footnote1}
In an arbitrary two-level quantum system there always exists a temperature value that parametrizes any mixed quantum state. This allows to keep for such a systems the expression of the exchange energy characteristic function $G(u)$ as a function of $u$, complex number, without necessarily making the usual substitution $u=i\beta$ with $\beta$ reference inverse temperature. That substitution, however, becomes relevant in proving the validity of the fluctuation theorem, where $\beta$ has to be taken equal to the inverse temperature of the initial thermal state.
\end{thebibliography}
\end{document}